\documentclass[conference]{IEEEtran}
\IEEEoverridecommandlockouts
\usepackage{cite}
\usepackage{amsmath,amssymb,amsfonts}
\usepackage{algorithmic}
\usepackage{graphicx}
\usepackage{textcomp}
\usepackage{xcolor}
\usepackage{algorithm,algorithmic}
\usepackage{url}
\usepackage{hyperref}

\def\BibTeX{{\rm B\kern-.05em{\sc i\kern-.025em b}\kern-.08em
    T\kern-.1667em\lower.7ex\hbox{E}\kern-.125emX}}
\begin{document}

\title{Metric-agnostic Learning-to-Rank via Boosting and Rank Approximation\\

}

\author{\IEEEauthorblockN{Camilo Gomez}
\IEEEauthorblockA{\textit{Dept. of Statistics \& Data Science} \\
\textit{University of Central Florida}\\
Orlando, USA \\
camilo.gomez@ucf.edu}
\and
\IEEEauthorblockN{Pengyang Wang}
\IEEEauthorblockA{\textit{Department of CIS, SKL-IOTSC} \\
\textit{University of Macau}\\
Macao, China\\
pywang@um.edu.mo}
\and
\IEEEauthorblockN{Yanjie Fu*\thanks{© 2023 IEEE. This is the author's accepted manuscript of a work accepted for publication in IEEE ICDM 2023. The final version is available at:  \href{https://doi.org/10.1109/ICDM58522.2023.00121}{https://doi.org/10.1109/ICDM58522.2023.00121}}}
\IEEEauthorblockA{\textit{School of Computing and AI} \\
\textit{Arizona State University}\\
Tempe, USA \\
yanjie.fu@asu.edu}
}
\maketitle
\begin{abstract}
Learning-to-Rank (LTR) is a supervised machine learning approach that constructs models specifically designed to order a set of items or documents based on their relevance or importance to a given query or context. Despite significant success in real-world information retrieval systems, current LTR methods rely on one prefix ranking metric (e.g., such as Normalized Discounted Cumulative Gain (NDCG) or Mean Average Precision (MAP)) for optimizing the ranking objective function.
Such metric-dependent setting limits LTR methods from two perspectives: 
(1) non-differentiable problem: directly optimizing ranking functions over a given ranking metric is inherently non-smooth, making the training process unstable and inefficient; 
(2) limited ranking utility: optimizing over one single metric makes it difficult to generalize well to other ranking metrics of interest. 
To address the above issues, we propose a novel listwise LTR framework for efficient and generalizable ranking purpose. 
Specifically, we propose a new differentiable ranking loss that combines a smooth approximation to the ranking operator with the average mean square loss per query. 
Then, we adapt gradient-boosting machines to minimize our proposed loss with respect to each list, a novel contribution. 
Finally, extensive experimental results confirm that our method outperforms the current state-of-the-art in information retrieval measures with similar efficiency. 

\end{abstract}

\section{Introduction}

In the realm of machine learning and information retrieval, ranking is a pivotal process that organizes a collection of items or documents in an order that reflects their relevance or significance to a specific query or context. This process is integral to a wide array of machine learning applications, encompassing search engines, recommendation systems, and natural language processing\cite{li2014learning}. 
A specialized branch of this discipline, known as Learning-to-Rank (LTR), employs a supervised machine learning approach to build models specifically designed to tackle ranking tasks. The learning-to-rank paradigm operates by training a model on a dataset composed of queries, each paired with a list of items. Each item within these lists is marked with a relevance grade, providing a measure of its pertinence to the associated query. This structured approach allows the model to learn the intricate relationships between queries and their relevant items, thereby enabling it to effectively rank new, unseen queries. 

Existing LTR literature can be categorized into three groups: 
(1) pointwise LTR, where each individual item in the training set is treated as an independent instance and is assigned a real-valued score or a class label indicating its degree of relevance to a query. The main drawback of pointwise LTR is ignoring the relative ordering or ranking of items within the same query, leading to suboptimal performance.
(2) pairwise LTR, that considers pairs of items in the training set and aims to learn a model that correctly orders each pair based on their relative relevance to a query. 
The limitation of pairwise LTR is neglecting the overall ranking structure among all items associated with a query, resulting in inconsistencies in the final ranking output.
(3) listwise LTR, which addresses the limitations of the above pointwise and pairwise LTR methods by treating the complete set of items linked to a query as a single entity during the training process, aiming to learn a model that optimizes the comprehensive ranking order of these items. 

Despite demonstrating notable performance, listwise learning-to-rank methods hinge on the optimization of specific ranking metrics, such as Normalized Discounted Cumulative Gain (NDCG) or Mean Average Precision (MAP). 
This reliance on metric-dependent learning criteria can give rise to two significant issues: 

\noindent \textbf{(1) non-differentiable problem}.  
To derive values for these ranking metrics, a ranking or sorting operation must be executed. 
However, these functions are inherently non-smooth, which poses a significant challenge when attempting to optimize them directly using gradient-based methods.
To address this issue, the optimization of a continuous approximation of these metrics has been suggested. For example, SoftNDCG \cite{r7} provides a smooth, differentiable approximation to the NDCG, which can then be minimized using gradient descent. 
Nevertheless, these continuous approximations have been found to be computationally intensive, rendering them impractical for use in large-scale production systems. Specifically, SoftNDCG and ApproxAP \cite{r8} require $O(n^3)$ and $O(n^2)$ computations respectively to achieve these ranking approximations or ``soft'' ranks.
Recently, an Optimal Transport (OT)-based soft ranking operator \cite{r9} has been introduced, which achieves soft ranking in $O(T nm)$, where $T$ is the number of Sinkhorn \cite{r10} iterations and $m$ and $n$ are hyperparameters. 
The first $O(nlogn)$ soft ranking with $O(n)$ differentiability has also been developed recently \cite{r11}. However, its performance in the context of Learning-to-Rank for Information Retrieval benchmark datasets (LETOR) remains to be evaluated, we do so in this paper. 

\noindent \textbf{(2) Limited ranking utility}. 
The information retrieval community has traditionally operated under the premise that optimizing evaluation metrics, such as NDCG or MAP, directly correlates with enhanced performance. 
However, recent studies have begun to challenge this assumption, suggesting that these metrics may intentionally serve as information bottlenecks and provide limited utility to user-centric applications \cite{r15} \cite{jiang2016correlation}.
The need for evaluating ranking utility over a large spectrum of increasingly complex metrics has in turn increased the need for metric-agnostic algorithms \cite{metricagnostic}. 
This raises a pertinent question: How can we develop a versatile ranker that is not constrained by specific evaluation metrics, yet still delivers performance on par with the current state-of-the-art methods when evaluated using commonly used metrics? 
The pursuit of such a general-purpose ranker would significantly enhance the flexibility and applicability of ranking algorithms in various contexts.

To address the above issues, in this paper, we propose a new performant metric-agnostic LTR framework for information retrieval systems. Our contributions and novelty can be summarized as follows: 
\begin{enumerate}
\item	We propose a novel differentiable loss function specifically designed for the ranking task. Our approach involves utilizing the average mean square loss between the true and predicted approximate ranks across all lists, thereby providing a genuine listwise measure. Specifically, we use ``fast-soft-ranking'' \cite{r11} as a building block for computing rank approximations. Our proposed measure is both continuous and differentiable, which are desirable properties for optimization.
\item To further enhance the learning process, we weigh these gradients by the derivatives of rank approximations. This strategy serves two key purposes. First, it eliminates the dependency on the evaluation metric, thereby increasing the flexibility of the ranking task. Second, it provides valuable ranking gradient information to the learners, which can guide the learning process more effectively.
The ultimate goal of this approach is to minimize the loss function with respect to the learnable ranking function. 
\item	To validate the performance and efficiency of our proposed approach, we conducted a comprehensive series of experiments using real-world data sourced from major commercial search engines. The results demonstrate that our approach consistently outperforms the state-of-the-art LambdaMART \cite{r12} across a variety of information retrieval metrics.
\end{enumerate}

The remainder of the paper is organized as follows. We briefly introduce the information retrieval ranking problem formally and notation used throughout the paper in section 2. In section 3, we present our algorithm SoftRankGBM. We describe in detail the modifications to gradient boosting machines (GBM) and its integration with a custom new differentiable loss function. Finally, in section 4, we present the our experiments.
\section{Preliminaries}

Several machine learning frameworks for handling the LTR problem have been proposed: pointwise, pairwise, and listwise. In information retrieval, ranking these objects depends on a context, such as a user query. Therefore, the aim is to optimize an objective function over all contexts (e.g., a list of queries). Because the objective depends on the rankings within each list, listwise methods have yielded better results in practice \cite{r17}. In this section, we formally present the listwise ranking problem and summarize the notation presented in Table I. 


\subsection{Listwise Learning-to-Rank}

The input space contains $K$ lists (queries) each of size $n^{(i)}$. Each list indexed by $(i)$ contains $n^{(i)}$  items denoted $\textbf{X}^{(i)} = (\textbf{x}_{1}^{(i)},\textbf{x}_{2}^{(i)},\ldots,\textbf{x}_{n^{(i)}}^{(i)} ) $, where $\textbf{x}_{j}^{(i)}$ denotes the $j$'th featurized item, usually documents. In output space, each list is associated with a ground truth $\textbf{y}^{(i)} = (y_{1}^{(i)}, y_{2}^{(i)},\ldots,y_{n^{(i)}}^{(i)} ) $. Normally, $y_{j}^{(i)}$  are the relevance labels. These ordinal labels form a permutation denoted $\pi_{y}^{(i)}$. Lastly, the training set $D$ contains all $K$ lists such that  $D =   \{ \textbf{X}^{(i)}, \textbf{y}^{(i)}\}_{ i=1}^{K} $.

The goal is to learn a hypothesis function that operates on a set of items and that it predicts an equivalent permutation as the ground truth. The perfect prediction, would be $h(\textbf{X}^{(i)})  = \pi_{y}^{(i)}$. Therefore, the listwise objective for this machine learning task aims to minimize the following empirical loss:

\begin{equation}
\min_{h} \frac{1}{K} \sum_{i}^{K}L(\pi_{y}^{(i)}, h(\textbf{X}^{(i)}))\label{empirical_loss}
\end{equation}

The learning system outputs a ranker $h(.)$. This ranker generates the predicted ranks for a new list of items $\textbf{X}^{(K + 1)}$. At test time, the ranker scores new lists individually. If predictions for multiple lists are required, the lists are scored separately and the evaluation metrics are aggregated (usually averaged) across all new lists.

\begin{table}[htbp]
\caption{Summary of Notation}
\begin{center}
\begin{tabular}{ll}
\hline
\textbf{{Variable}}& \textbf{Description} \\ \hline
$K$ & Total number of lists (queries) in the dataset. \\
$n^{(i)}$ & Number of documents in query $i$.\\
$\textbf{x}_j^{(i)}$ & Featurized document document $j$ of query $i$.\\
$\textbf{X}^{(i)}$ & List of all featurized documents in query $i$.\\
$\textbf{y}^{(i)} $ & List of relevance labels in query $i$. \\
$\textbf{X}^{(K+1)}$ & Test query.\\
$h(.)$ & Ranking function (ranker).\\
$\pi_y^{(i)}$ & Permutation formed by the labels of query $i$. \\
$NDCG@k $ & Mean NDCG truncated at k over all queries. \\
$MAP@k $ & Mean MAP truncated at k over all queries. \\
\hline

\end{tabular}
\label{notation}
\end{center}
\end{table}
\section{Methodology}

\subsection{Overview of the Proposed Ranking System}
Ranking evaluation metrics sort the predictions and ground truth labels to assess a ranker’s predictive quality. It is imperative, then, that our ranker $h$ produces high-quality rankings to improve the evaluation metrics. Because, in practice, the relevance labels are the true document ranks (with ties), we start constructing a ranker with a specific form. We propose to have a ranking operator and scoring function composition, i.e., $h(\textbf{X}^{(i)})= rank \circ f(\textbf{X}^{(i)})$. This way, we can write down an expression for the derivative of the cost \eqref{empirical_loss} using the chain rule as follows:

\begin{equation} \frac{\partial L}{\partial h}= \frac{\partial L }{\partial rank} \frac{\partial rank}{\partial f} \label{cost_derivative}
\end{equation}

Computing $\frac{\partial rank}{\partial f}$ is challenging because the ranking operation is non-smooth, as mentioned in Section 1. In the following two subsections, we describe how to train and output a general ranker that minimizes the listwise ranking loss from an approximation to the gradient in \eqref{cost_derivative}. We now describe how to construct our metric-agnostic method for ranking:

\begin{itemize}
    \item \textbf{(1)} First, we propose a differentiable loss function based on the average mean squared error (MSE) and an approximation to the ranking operator known as fast-soft-rank \cite{r11} denoted $r_{Q \epsilon}(.)$. We term our new loss \textbf{SoftRankMSE}.
    \item  \textbf{(2)} Then we modify gradient boosting to learn from these functional gradients while maintaining the listwise structure during training. For this reason we name our method \textbf{SoftRankGBM}, which stands for Soft Ranking Gradient Boosting Machines.
\end{itemize}

\subsection{A New Differentiable Ranking Loss}
There are many approximations to the ranking operator \cite{r7} \cite{r8} \cite{r9} \cite{r11}. These approximations seek to provide a continuous and differentiable proxy for the non-smooth ranking operator. However, only \cite{r11} has desirable properties that make it particularly useful for large-scale LTR applications. Among these properties are differentiability in $O(n)$ time complexity, order-preserving, and efficiency as it computes rankings in $O(nlogn)$ time complexity. Due to its properties, the authors coined this operator the ``fast-soft-ranking'' operator.

\subsubsection{Rank Approximation}
Our method builds on an approximation of the ranking operation. We use ``fast-soft-ranking'' as a building block in our algorithm, so we introduce it briefly.
First, let $\sigma$ denote a permutation of $n$ integers, and $\Sigma$ denote the set of all $n!$ permutations. Then, their idea was to first cast the ranking problem as a linear program over the convex set of all permutations, i.e., the permutahedron: 
\begin{equation}
\psi (\theta) := conv(\{ \theta_{\sigma} : \sigma \in \Sigma \} \subset \mathbb{R}^n) \label{permutahedron}
\end{equation}

Then using quadratic regularization $Q(.) = \frac{1}{2} ||.||^2 $ and a tuning parameter $\epsilon$, the soft ranking problem can be defined as follows, let $\theta \in \mathbb{R}^n$  and $\rho := (n,n-1,\ldots,1)$ Where $0 < \epsilon < \infty $, which controls the approximation to the ``hard'' ranks.

\begin{equation}
r_{Q \epsilon}(\theta) = \arg \max_{y \in \psi(\rho)}\langle y, - \theta / \epsilon  \rangle - \frac{1}{2} ||\rho||^2 \label{soft_rank}
\end{equation}

Lastly, casting as an isotonic optimization problem using simple chain constraints is used to obtain fast computation and differentiability.  We refer readers to \cite{r11} for details and mathematical proofs.

\subsubsection{Soft-Rank-MSE Loss}
Our idea stems from the fact the mean ranking quality of all the lists, such as the one measured by the mean NDCG or  MAP can be improved by increasing the score of individual lists. Naturally, the mean score will be larger if each list has a better score. Therefore, we focus on improving the quality of lists individually. For that reason, we use the mean squared error (MSE) loss between the ground truth and predicted ranks. This average-listwise-MSE provides differentiability to the first term $\frac{\partial L}{\partial rank}$
 in \eqref{cost_derivative}. To make term $\frac{\partial rank}{\partial f}$ differentiable, we use an approximation to the ranking operator as a drop-in replacement in $h$ and to compute the soft ranks for the relevance labels  $\textbf{ y}^{(i)}$. The relevance labels become $R_{Q}^{(i)} = r_{Q \epsilon}( \textbf{ y}^{(i)})$. Because $r_{Q \epsilon} $ is order preserving as seen in [6, Property 2], the drop in replacement does not affect the evaluation metrics. The scoring function becomes $\hat{R}_Q^{(i)}(\textbf{X}^{(i)} ) = h(\textbf{X}^{(i)})= r_{Q \epsilon} \circ f(\textbf{X}^{(i)})$ providing differentiability [13, Property 1] to the following loss:

\begin{equation}
L = \frac{1}{K} \sum_{i=1}^{K} \frac{1}{2n^{(i)}} \|R_{Q}^{(i)} - \hat{R}_Q^{(i)}(\textbf{X}^{(i)} ) \|^{2}\label{MSE_Soft_Rank}
\end{equation}

\subsection{Integrating Boosting into Differentiable Ranking}

\subsubsection{Boosting}
SoftRankGBM uses Gradient Boosting Regression Trees (GBRT), a tree-based gradient boosting machine (GBM), to learn the scoring function $f$. We briefly introduce tree boosting. The GBRT (or MART) algorithm learns a scoring function as an ensemble of sequential regression trees based on gradient boosting and combines the output of each tree in a linear additive structure. At each iteration, a tree is learned to minimize the residuals of the previous iteration.  In other words, out of all possible learnable functions $F$, which regression tree minimizes the loss. For that reason, GBRT is viewed as performing gradient descent in functional space \cite{r18}.
\subsubsection{Integration}
We now adapt the GBRT, which acts on individual samples to work for listwise LTR to minimize the SoftRankGBM loss. For this step, its important to note the training happens at a listwise level but scoring happens one new list at a time. Therefore, we make the learner implicitly aware of the listwise-structure information within the functional gradients, but explicitly train on individual samples. At each iteration $t$, we evaluate the ensemble built so far (i.e., $f_{t-1}$) composed with the soft rank operator to generate the current model's predicted ranks for all $i=1,\ldots,K$ lists in parallel:

  \begin{equation}
 \hat{R}_{Q t-1}^{(i)}(\textbf{X}^{(i)}) = r_{Q \epsilon} \circ f_{t-1}(\textbf{X}^{(i)})  \label{new_evaluation}
\end{equation}
We can use these predicted ranks to compute the loss \eqref{MSE_Soft_Rank} and its gradient. However, we do so evaluating using only the current list's $n^{(i)}$ entries.
  \begin{equation}
\frac{\partial L(R^{(i)}_{Q}, \hat{R}_{Qt-1}^{(i)})}{\partial \hat{R}_{Qt-1}^{(i)}} \label{partial_listwise}
\end{equation}
Resulting in $K$ listwise partial derivatives of the loss with respect to each query list at iteration $t$. Lastly, we concatenate these and treat them as the new residuals by letting the rest of the learning continue as in regular GBRT.

\subsubsection{Algorithm}
In this section we present the complete SoftRankGBM approach, which we summarize in Algorithm 1. SoftRankGBM  has four parameters, namely, the number of iterations $T$, the learning rate $\gamma$, the number of leaf nodes per tree $L$ and the approximation parameter to the rank operator $\epsilon$. We assume there are $K$ lists and $N = \sum_{i=1}^{K}n^{(i)}$ total documents. The initialization step consists of predicting a constant for  every list $f_0 = c$. For each boosting iteration $T$, in step 4 the soft-ranks of every list are computed in parallel as in \eqref{new_evaluation}. Once the loss and the gradient are computed in step 5, all partial derivatives are stacked into a vector containing all the negative $K$ listwise gradients. Where the negative partial derivative of the loss with respect to list $i$ at time $t$ are denoted $\nabla_{t}^{(i)}$.

The resulting vector $\textbf{q}_t = vec(\nabla_{t}^{(1)}, \nabla_{t}^{(2)}, \ldots \nabla_{t}^{(K)})$ is of size $N$. Step 8 fits a new regression tree on the new dataset $\{\textbf{X}, \textbf{q}_t \}$, where $\textbf{X}^\intercal= [X^{(1)} X^{(2)} \dots X^{(K)}]^\intercal$ are all the document features stacked on a matrix. A small nuance and novelty of our approach is that during the tree-fitting stage, in step 8, the feature-target relationship is learned across queries and for all documents. Experimentally, we see the performance benefits to this approach.

\begin{algorithm}
 \caption{SoftRankGBM}
 \begin{algorithmic}[1]
 \renewcommand{\algorithmicrequire}{\textbf{Input:}}
 \renewcommand{\algorithmicensure}{\textbf{Output:}}
 \REQUIRE Training set $D$; the number of iterations $T$; the number of leaf nodes $L$; the learning rate $\gamma$; the soft-ranking parameter $\epsilon$
 \ENSURE  prediction $f(\textbf{X})$ for an instance $\textbf{X}$
  \STATE Initialize the function $f_0 = 0$
  \FOR {$t = 1$ to $T$}
  \FOR {$i=1$ to $K$ \textbf{parallel}} 

  \STATE Compute the ranks with respect to each list $i$, 
    \begin{equation}
    \hat{R}^{(i)}_{Qt-1}(\textbf{X}^{(i)}) = r_{Q \epsilon}(f_{t-1}(\textbf{X}^{(i)}))
  \label{predicted_soft_ranks_in_algo}
  \end{equation}
 \STATE Calculate the residual vector $\nabla^{(i)}_{t}$ as the partial derivatives of the expected loss function $L(R^{(i)}_{Q}, \hat{R}_{Q}^{(i)})$ at each point of each list: 
  \begin{equation}
\nabla^{(i)}_{t} = 
- \left( \frac{\partial L(R^{(i)}_{Q}, \hat{R}_{Qt-1}^{(i)})}{\partial \hat{R}_{Qt-1}^{(i)}} \right)\label{negative_gradients}
\end{equation}

  \ENDFOR
  \STATE Stack all residuals into a vector
    \begin{equation}
\textbf{q}_t = vec(\nabla^{(i)}_t), i=1,\ldots,K\label{pseudo_residuals}
\end{equation}
  \STATE Train base model $h_t(\textbf{X})$ on a new dataset with residuals $\{\textbf{X}, \textbf{q}_t \}$
  \STATE Update function $f_t(\textbf{X}) = f_{t-1}(\textbf{X}) + \gamma_t h_t(\textbf{X})$
  \ENDFOR

 \STATE The resulting function after $T$ iterations is
  \begin{equation}
 f_T(\textbf{X}) = \sum_{t=1}^{T}\gamma h_t(\textbf{X}) = f_{T-1}(\textbf{X}) + \gamma h_{T} (\textbf{X})  \label{Ranker}
\end{equation}
 \end{algorithmic} 
 \end{algorithm}
\section{Experiments}
In this work, we performed extensive experiments on two benchmark datasets, “C14!“ and “Web10k”. Both datasets from major commercial search engines (Yahoo and Bing, respectively) have been used extensively in the information retrieval literature to evaluate ranking performance due to their popularity and public availability. We also compared SoftRankGBM with the state-of-the-art LambdaMART and Adarank as baselines. 

\subsection{Questions to Study in Experiments}
In the following subsections, we aim to answer the following research questions:

\begin{itemize}

\item	\textbf{Q1:} What is the performance of SoftRankGBM in the ranking task compared to the state-of-the-art?
\item	\textbf{Q2:} How does the efficiency of training compare between SoftRankGBM and state-of-the-art implementations of the baseline methods?
\item	\textbf{Q3:} How critical is each component of our proposed ranking system SoftRankGBM?

\end{itemize}

\subsection{Data Description}
In our experiments, we evaluate the model performance on two different benchmark datasets. These datasets have been made publicly available by major search engines for evaluating the performance of learning-to-rank methods. Due to their popularity, the term LETOR (Learning to Rank for Information Retrieval) datasets was coined. The statistics of these datasets are summarized in Table III, and details are shown as follows:

\begin{table*}[htbp]
\caption{Benchmark on LETOR datasets}
\begin{center}
\begin{tabular}{|c|c|c|c|c|c|c|c|c|}
\hline
\textbf{}&\multicolumn{4}{|c|}{\textbf{WEB10k}} &\multicolumn{4}{|c|}{\textbf{Yahoo C14!}}  \\ \cline{2-9} 
~ & \textbf{\textit{ndcg@1}} & \textbf{\textit{ndcg@10}} & \textbf{\textit{map@1}} & \textbf{\textit{map@10}} & \textbf{\textit{ndcg@1}} & \textbf{\textit{ndcg@10}}  & \textbf{\textit{map@1}} & \textbf{\textit{map@10}}  \\ \hline
SoftRankGBM (ours) &  \textbf{0.4813} & \textbf{0.5004} & \textbf{0.8100} & \textbf{0.6433}   & \textbf{0.7241} & \textbf{0.7930} & \textbf{0.9155} & \textbf{0.8728}  \\ \hline
LambdaMART (Lightgbm) &  0.4680 & 0.4899 & 0.7815 & 0.6250  & 0.7197  &  0.7923 & 0.9102 & 0.8690   \\  \hline
LambdaMART (XGBoost)$^{\mathrm{a}}$ &   0.4522 & 0.4796 & - & -  & 0.7071 & 0.7698  & - & -   \\ \hline
LambdaMART (RankLib)  &   0.4281 & 0.4486 & 0.7220  & 0.6377  & 0.6860  & 0.7478  & 0.8783 & 0.8102   \\ \hline
Adarank (RankLib) &   0.2626 & 0.3350 & 0.6880 & 0.5989 & 0.6438 & 0.7067  & 0.8430 & 0.8135   \\ \hline
\multicolumn{5}{l}{$^{\mathrm{a}}$XGBoost's MAP only works with binary labels.}
\end{tabular}

\label{tab1}
\end{center}
\end{table*}

\begin{itemize}
    \item \textbf{LETOR WEB10K Data}. The WEB10K dataset has been widely used as a benchmark in evaluating the performance of LTR algorithms. This dataset has been partitioned into several parts, which we use {S1, S2, S3} for training and {S5} for validation. These partitions contain a total of 964,933 URLS (candidate documents) and 8,000 queries. The relevance judgments are obtained from Microsoft's Bing, a commercial search engine, which take 5 values from 0 (irrelevant) to 4 (perfectly relevant) \cite{r29}.
    \item \textbf{Yahoo C14!} Data. The C14! data is another benchmark dataset for LTR performance validation. This dataset has been partitioned into several parts of which we use Set1. This partition containing a total of 544,217 URLs (candidate documents) and 22,938 queries \cite{chapelle2011yahoo}.

\end{itemize}

\begin{table}[htbp]
\caption{Statistics of the two datasets used in our LTR experiements}

\begin{center}

    \begin{tabular}{|c|c|c|c|c|}
    \hline
        ~ &\multicolumn{2}{|c|}{\textbf{YAHOO}} & \multicolumn{2}{|c|}{\textbf{WEB10k}} \\ \cline{2-5} 
        ~ & Train & Valid. & Train & Valid. \\ \hline
        Queries & 19,944 & 2,994 & 6,000 & 2,000 \\ \hline
        URLs (Docs) & 473,134 & 71,083 & 723,412 & 241,521 \\ \hline
        Features & \multicolumn{2}{|c|}{700} & \multicolumn{2}{|c|}{136} \\ \hline
    \end{tabular}
\label{tab2}
\end{center}
\end{table}

\subsection{Evaluation Metrics}
For benchmarks we use the NDCG and MAP \cite{r28} and their truncated versions the $k$ denoted NDCG@k and MAP@k, which are the most popular metrics in IR. Specifically, because the top $k$ are of particular interest to commercial search
engines in information retrieval, we used the following levels $k = [1, 10]$.

\subsection{Baseline Methods for Comparison}
For this comparison, we test with the current state-of-the-art LambdaMART and Adarank \cite{r23} as baseline models. However, this method has different implementations with varying amounts of performance. In the past, many authors have compared their methods to lesser-performant implementations, such as RankLib, and claimed state-of-the-art results \cite{r14}. For this reason, we compare amongst all popular LambdaMART implementations, i.e., LightGBM~\cite{ke2017lightgbm}, XGBoost~\cite{chen2016xgboost}, and RankLib \cite{dang2013lemur}. 


\subsection{Reproducibility and Parameter Settings}
 All methods are executed for $1,000$ boosting iterations with a learning rate $\gamma=0.1$, soft-ranking parameter $\epsilon = 0.01$, and number of leaf nodes $L = 255$. All the other parameters are set to their default in LightGBM, XGboost and RankLib respectively.

\subsection{Performance Comparison (Q1)}

We evaluated the performance of all compared algorithms on two LETOR benchmark datasets and reported their evaluation results in Table II. We highlighted the best performant score per column (higher is better). (1) We can observe that SoftRankGBM consistently outperforms all implementations of LambdaMART as measured by both the MAP and NDCG metrics at different truncation levels. (2)  We notice that LambdaMART's testing accuracy stops increasing after around 100 iterations (Fig. 1 left). However, although SoftRankGBM takes more iterations to reach its most performant model, our method's testing accuracy keeps increasing with more iterations. 

\begin{table*}[htbp]
\caption{Ablation study}
\begin{center}
\begin{tabular}{|c|c|c|c|c|c|c|c|c|}
\hline
~ &\multicolumn{4}{|c|}{\textbf{WEB10K}} &\multicolumn{4}{|c|}{\textbf{Yahoo C14!}} \\ \cline{2-9} 
~ & \textbf{\textit{ndcg@1}}& \textbf{\textit{ndcg@10}}& \textbf{\textit{map@1}} & \textbf{\textit{map@10}} & \textbf{\textit{ndcg@1}}& \textbf{\textit{ndcg@10}}& \textbf{\textit{map@1}} & \textbf{\textit{map@10}}\\ \hline
GBRT & 0.4701 & 0.4908 & 0.7490 & 0.5854 & 0.7239 & \textbf{0.7934} & 0.9108 & 0.8696 \\  \hline
GBRT + SoftRankMSE & 0.4727 & 0.4944 & \textbf{0.8135} & \textbf{0.6440} & 0.7207 & 0.7883 & 0.9119 & 0.8670\\ \hline
GBRT (Listwise)& 0.3788 &0.4130 & 0.7335  & 0.5620 & 0.6611 
& 0.7405 & 0.8829 & 0.8338\\ \hline
SoftRankGBM & \textbf{0.4813} & \textbf{0.5004} &  0.8100  & 0.6433 & \textbf{0.7241}  & 0.7930 & \textbf{0.9155} & \textbf{0.8728}\\ \hline

\end{tabular}
\label{tab3}
\end{center}
\end{table*}

\begin{figure*}[htbp]
    \centering
    \includegraphics[width=\linewidth]{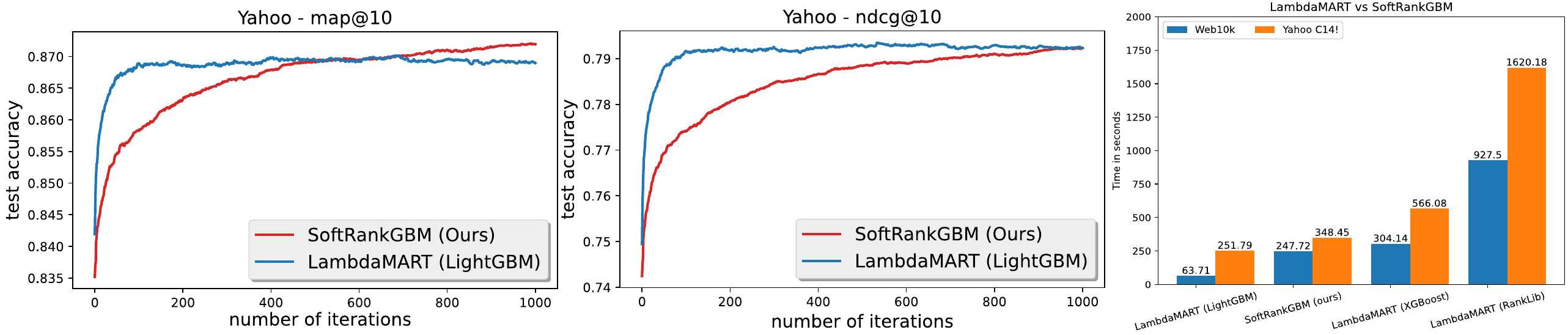}
    \caption{Learning curves of SoftRankGBM (ours) and LambdaMART (left and center). Training times (right)}
\end{figure*}

\subsection{Train-time efficiency (Q2)} 
We compared the training time of SoftRankGBM against the train time of LambdaMART. However, there are multiple implementations of LambdaMART with various degrees of performance and efficiency. Our method had the second best performance in the comparison (Fig. 1 right) as it is only outperformed by the LightGBM implementation. However, for the two other implementations, namely, XGBoost and Ranklib, our method is competitively better.

\subsection{Model Ablation Study (Q3)} 
In addition to state-of-the-art comparisons, we were also interested in better understanding the proposed approach and evaluate its key components. Particularly, how crucial is each component to the adaptation from gradient boosted regression trees to SoftRankGBM? Hence, in our evaluation, we consider the following three variants:

\begin{itemize}
    \item \textbf{GBRT.} We use the pointwise ranking approach and treat the problem in the traditional regression framework using Gradiant Bosted Regression Trees. We ignore the query-document dependency and treat each document as an individual sample using the squared loss.
    
    \item \textbf{GBRT with SoftRankMSE loss.} We use the pointwise ranking approach but this time we use the information of the soft rank gradients used in SoftRankGBM. However, we do not account for the query-document dependency, we still individual samples for training.
    
    \item \textbf{Listwise GBRT.} We adapt GBRT from a pointwise to a listwise method. Here we use the sum of squared losses across all queries, however only we do not use the SoftRankGBM gradients.
\end{itemize}

Our proposed variant is superior to other variants 5 out of 8 times (Table IV). We notice that another 2 of these 8 times, our complete method was outperformed by a variant of GBRT when combined with our SoftRankMSE loss. This suggests that minimizing our SoftRankMSE loss is an effective method to improve ranking metrics (7 out of 8 trials). However, for listwise problems, that is, when items have a nested structure, such as in commercial search engines, our complete method SoftRankGBM yields more desirable results.

\section{Conclusions}
We proposed SoftRankGBM, a novel metric-agnostic learning to rank method that leverages the average mean square error and a smooth approximation to the ranking operator \cite{r11} as building blocks. We then modified gradient boosting to optimize our differentiable loss function. This allows us to obtain a listwise metric-agnostic performant ranker. In particular, we measured our approach using commonly used ranking metrics for information retrieval systems such as NDCG and MAP.
We conducted experiments on popular and publicly available LTR datasets for information retrieval (LETOR), and observed that SoftRankGBM was superior on $8/8$ metric-truncation level combinations across these datasets. Furthermore, we conducted ablation studies to evaluate the individual contribution of each component of our framework separately. These experiments suggest our full adaptation, that is, using gradient boosting to optimize our proposed metric-agnostic loss, outperforms the other variants.
\section*{Acknowledgment}

This work is partially supported by NSF IIS-2152030, IIS-2045567, IIS-2006889, IIS-2040950.


\bibliographystyle{IEEEtran}
\bibliography{IEEEabrv,references}


\end{document}